\documentclass[10pt,aps,pra,twocolumn,showpacs,nofootinbib,longbibliography,notitlepage,superscriptaddress]{revtex4-2}

\usepackage{etex}
\usepackage{amsmath,amssymb,amsthm}

\usepackage{graphicx}
\usepackage{cancel}
\usepackage[scr=rsfs]{mathalpha} 
\usepackage{subcaption}

\usepackage[colorlinks=true,citecolor=blue,urlcolor=blue]{hyperref}
\usepackage{times,txfonts}
\usepackage{physics}
\usepackage{braket}
\usepackage{blkarray}
\usepackage{mathtools}
\usepackage{latexsym}
\usepackage{tabularx, booktabs}
\usepackage{graphics,epstopdf}
\usepackage{float}
\usepackage{amsfonts}
\usepackage{color,soul}

\newcommand{\be}{\begin{equation}}
        \newcommand{\ee}{\end{equation}}
\newcommand{\ba}{\begin{eqnarray}}
	\newcommand{\ea}{\end{eqnarray}}

\newcommand{\half}{\frac{1}{2}}

\newcommand{\etal}{{\it{et al. }}}
\newcommand{\pr}{\operatorname{Pr}\{\operatorname{success}\}}

\newcommand{\pcl}{\operatorname{P}^{\textsl{cl}}}
\newcommand{\pqm}{\operatorname{P}^{\textsl{qm}}}
\newcommand{\Ccl}{\mathscr{C}_{n}^{\textsl{cl}}}
\newcommand{\Cqm}{\mathscr{C}_{n}^{\textsl{qm}}}

\begin{document}

\title{Intraparticle entanglement-based Random Access Code protocols: Contextuality-enabled quantum advantage and implications}



\author{Nilaj Saha}
\email{nilajsaha29@gmail.com}
\affiliation{Indian Institute of Science Education \& Research (IISER), Mohali, SAS Nagar, Punjab 140306, India}

\author{Sumit Mukherjee}
\affiliation{Korea Research Institute of Standards \& Science (KRISS), Gajeong-ro, Daejeon 34113, Republic of Korea}
\affiliation{Department of Physics of Complex Systems, S. N. Bose National Center for Basic Sciences, Block JD, Sector III, Salt Lake, Kolkata, West Bengal 700106, India}

\author{Dipankar Home}
\affiliation{Raman Research Institute (RRI), C. V. Raman Avenue, Sadashivanagar, Bengaluru, Karnataka 560080, India}
\affiliation{Bose Institute, Kolkata, West Bengal 700091, India}




\begin{abstract}
\vspace{0.5mm}
We provide the first explicit identification and quantitative characterization of the physical origin of the quantum advantage in the Random Access Code (RAC) protocol. This is achieved by formulating the protocol in terms of intraparticle entanglement between co-measurable degrees of freedom of a single particle and establishing a fundamental correspondence between the protocol's success probability and the underlying resource powering it. For this purpose, we use a relevant Bell-type inequality derived from the assumption of noncontextuality of measurement outcomes. The formulated analysis reveals that the magnitude of quantum-mechanical violation of this inequality, signifying a form of quantum contextuality, is quantitatively commensurate with the ``quantum enhancement" of success probability in any intraparticle entanglement-assisted $n$-bit RAC protocol. In particular, the maximal success probability achievable in a quantum $n \mapsto 1$ RAC protocol corresponds to the maximal quantum violation of the relevant Bell-type inequality. Our framework not only demonstrates how quantum contextuality entailed by intraparticle entanglement serves as an effective resource for enhancing RAC performance, but also offers a significant operational advantage: the proposed scheme is readily implementable in a single-particle interferometric setup requiring coherence preservation only for a single particle, rather than between spatially separated entangled systems.
\end{abstract}
\maketitle


\section{Introduction}
A hallmark of the classical worldview is the notion of noncontextuality which implies that the properties of an individual system can be defined independently of their own measurement, and irrespective of what other measurements are made. Incompatibility between this premise and quantum mechanics has been extensively studied from various perspectives, originating from the studies by Gleason~\cite{gleason57} and Kochen-Specker~\cite{KSpeker68}, eventually leading to a number of experiments~\cite{mich00,simon2000,hasegawa03,contextexp1,contextexp2,contextexp3,contextexp4,contextexp5,contextexp6}, establishing this incompatibility as an ineluctable feature of both quantum mechanics and nature (for a recent review, see Budroni \etal~\cite{contextrev}). In contrast to the other fundamental quantum feature of nonlocality, which occurs essentially for spatially separated entangled particles, a single system is sufficient for evidencing the quantum feature of contextuality. The Kochen-Specker notion of measurement contextuality has subsequently been generalised~\cite{spekkens05,hazra24} beyond sharp measurements to arbitrary experimental procedures, as well as to arbitrary operational theories by going beyond standard quantum theory. Extensive studies have, in turn, led to a plethora of applications in a variety of areas related to communication games~\cite{hameedi,pan19}, state discrimination~\cite{schmid18,mukherjee22,flatt22}, and various other information theoretical tasks~\cite{spek09,magic,ghorai,saha19b,kumari2019,lostaglio20,gupta23}. \vspace{2mm}

Complementing the above range of studies, another approach was initiated~\cite{home1984,basu2001} using the notion of intraparticle entanglement as resource for exploring quantum contextuality. This concept considers entanglement between jointly measurable dynamical variables (such as path and spin/polarization) pertaining to disjoint Hilbert spaces of the same particle; in this context, the relevant nonseparable states are known as intraparticle entangled states. It was shown~\cite{home1984} that the Bell-type inequality, derived from the notion of noncontextuality applied to two such degrees of freedom of a single particle, is violated by the quantum mechanical predictions for an intraparticle entangled state. Further, this basic idea was formulated in a way making it amenable to an experimental test~\cite{basu2001}. Subsequently, such a test was realised using photon~\cite{mich00} and neutron~\cite{hasegawa03} interferometries. The conceptual significance alongside usefulness of intraparticle entanglement and its various features have been investigated by a range of studies~\cite{quantum_gates&circuit_1998,akpan09,swap,qrep,qt1,qkd2,qt2,qkd1,qkd3,mazzucchi21,adc}, comprehensively reviewed by S. Azzini \etal~\cite{azzini20}. \vspace{2mm}

Against this backdrop, we focus on examining an earlier unexplored question: To what extent quantum contextuality arising from intraparticle entanglement can serve as a resource in Random Access Code (RAC) protocols, and whether there exists a deeper fundamental connection between the success probability of such protocols and the underlying resource powering them? It is relevant to note here that for a suitably formulated communication task, Spekkens \etal had given a formal argument suggesting information theoretic advantages that can stem from single system based "preparation contextuality"~\cite{spek09}; such an argument was then extended for other communication protocols~\cite{hameedi}, as well as by using what has been called "universal quantum contextuality"~\cite{pan19}. On the other hand, in this paper, by considering our formulated intraparticle entanglement-based versions of the RAC protocol, we show how contextuality manifested through such intraparticle path-spin entanglement in an appropriate interferometric arrangement, can provide significant quantum enhancement of success probabilities. We further investigate how this quantum advantage is exactly related to the violation of suitably formulated noncontextuality-based Bell-CHSH-type inequalities. \vspace{2mm}

The present paper is organized as follows. In order to set our work in proper perspective, we begin with a broad recapitulation of the different versions of the RAC protocol and highlight their overall significance (Section \ref{sec:2}). Subsequently, we first focus on the classical 2-bit RAC protocol (Section \ref{sec:2a}) and proceed in three steps: (a) Considering all possible strategies, it is shown that the success probability attains optimal value for suitable "deterministic strategies". (b) The protocol is then appropriately formulated to apply the notion of noncontextuality. In this context, a Bell-type inequality is invoked, which is derivable from this assumption of noncontextuality. (c) The value of the Bell-type correlation variable occurring in this inequality is shown to be related to the classical success probability of the protocol, such that the upper bound of this noncontextuality-based inequality reproduces the optimal classical success probability mentioned earlier in (a). \vspace{2mm}

Next, switching our attention to the quantum scenario of the 2-bit RAC protocol, we proceed through the following two stages: (i) With respect to the framework mentioned above, we formulate the quantum scenario of the protocol by utilising an intraparticle path-spin entangled state as resource. For this purpose, a suitable Mach-Zehnder (MZ) type interferometric arrangement is considered. (ii) The information-theoretic quantum advantage of our formulated protocol is demonstrated by showing that the success probability of the protocol is enhanced compared to its classical counterpart, the amount of this enhancement been determined by the magnitude of the quantum mechanical violation of the Bell-type inequality in question. It is worth stressing that while the MZ-type interferometric arrangement based on intraparticle path-spin entanglement has been earlier discussed~\cite{basu2001, hasegawa03, Wagner2024_MZI_coh&context}, its use in harnessing intraparticle entanglement as resource for RAC protocols has remained hitherto uninvestigated. Also, note that, although our treatment is presented in terms of spin-$\frac{1}{2}$ particles (such as neutrons), it is equally applicable for photons by considering the polarization degree of freedom instead of spin, and by using polarizing and analyzing devices appropriate for photons in the setup. \vspace{2mm}

The above formulation is then extended to the 3-bit RAC protocol by suitably redefining the relevant measurement bases (Section \ref{sec:2b}). Notably, this extension can be achieved within the same setup and with the same path-spin intraparticle entangled state used for the 2-bit protocol. Here too, we formulate the noncontextuality-based Bell-type inequality relevant to the 3-bit RAC protocol such that it can then be invoked to obtain an appropriate upper bound for the classical success probability of the protocol. We show that this bound, too, can be surpassed in the quantum scenario by leveraging the violation of the considered Bell-type inequality, with the increase in success probability being directly commensurate with the quantum violation of the noncontextual bound. Finally, our approach is generalised (Section \ref{sec:2c}) by deriving an expression for the success probability of our intraparticle entanglement-assisted $n$-bit RAC protocol in terms of the violation of an appropriately defined noncontextuality-based Bell-type inequality whose form is analogous to Gisin's "elegant Bell inequalities"~\cite{gisin}. In the concluding Section \ref{sec:3}, the salient features and results of this work have been encapsulated, capped by indicating a number of directions for future studies.


\section{Different versions of the RAC protocol}
\label{sec:2}
Random Access Code (RAC) is a widely useful communication protocol which enables extracting information about the value of a randomly selected bit of a given sequence of bits from its substring communicated to a receiver. To put it more explicitly, RAC entails a sender (Alice) encoding an $n$-bit message into a shorter $m$-bit string, typically denoted by $n \stackrel{p}{\mapsto} m$ (where $n > m$), and sends it to Bob, who then attempts to decode the value of a randomly selected bit; here $p$ denotes the average success probability of such extraction of information. While this protocol is usable in the classical context~\cite{ambainis08}, the quantum formulation of this protocol can be realized in the following two ways: (a) Alice sends Bob quantum states encoding the desired $n$-bit string into an $m$-qubit message. Subsequently, Bob decodes this message by measuring the quantum state. We term this as prepare-measure Quantum Random Access Code or simply QRAC~\cite{stephen83}. (b) In the other scheme, instead of Alice sending qubits to Bob, Alice and Bob can share an entangled state with respect to which Alice makes a measurement to remotely prepare certain quantum state for Bob on which Bob makes measurements to finally decode the desired bit from the encoded bit string with the help of appropriately preshared classical bits. This is termed as Entanglement-Assisted Random Access Code (EARAC)~\cite{EARAC10}. \vspace{1mm}

Subsequent to the pioneering work on QRACs by Wiesner~\cite{{stephen83}}, extensive studies were made by Ambainis \etal in~\cite{{ambainis08}} comparing QRACs with their classical counterparts. Notably, they formulated and evaluated the success probabilities of the $n \mapsto 1$ QRACs (for $n \in \{2,3,4,5,6,9,15\}$) using numerical methods. Further work by Paw\l{}owski \& \ifmmode \dot{Z}\else \.{Z}\fi{}ukowski in~\cite{EARAC10} formulated EARACs and teased out its implications. Subsequently, the 2 $\mapsto$ 1 \& 3 $\mapsto$ 1 EARAC protocols were realised experimentally~\cite{wang19}. Over the years, RAC protocols have been having multifaceted applications in the areas related to network coding~\cite{app1,netcode}, locally decodable codes~\cite{app2}, semi-device independent random number generation \& certification~\cite{app6,app5}, quantum communication complexity~\cite{app13,app12}, dimension witnessing~\cite{app9}, self-testing~\cite{app10}, stochastic teleportation~\cite{app_teleport_2026}, and quantum key distribution~\cite{app11}. Further, ramifications of the RAC protocols for the use of computational learning method to perform quantum state tomography have been analysed~\cite{app3,app4}. From a foundational perspective, RAC has been analysed within the framework of generalized probability theory in the context of the Popescu-Rohrlich (PR) box example~\cite{app7,app8}. \vspace{1mm}

Now, an important point to stress here is that the two quantum formulations of RAC (QRAC \& EARAC) described earlier are equivalent in terms of the success probabilities of the protocols for $n \leq 3$, \textit{i.e.} for the 2-bit ($2 \mapsto 1$) and 3-bit ($3 \mapsto 1$) cases. To put it more precisely, the success probabilities for the 2-bit and the 3-bit EARACs are exactly the same as those of the corresponding QRACs~\cite{ambainis08,EARAC10}. Furthermore, note that an $n$-bit EARAC can be constructed for any $n$, whereas for $n \geq 4$, $n$-bit QRACs achieving a success probability greater than the trivial value of 0.5 do not exist without Shared Randomness (SR)~\cite{app1}. Even when QRACs are implemented using SR, they yield lesser success probabilities than the corresponding EARACs for all $n \geq 4$~\cite{EARAC10}. With this perspective, in the final part of our present work, where we discuss the generalisation of our obtained results for an $n$-bit scenario ($n \mapsto 1$), the concatenation scheme~\cite{EARAC10} is utilised. Note that while generalising for $n$-bits, considering this is in itself sufficient. This is because any bound which is applicable to the success probabilities of $n$-bit EARACs would be equally applicable to the success probabilities of the corresponding QRACs~\cite{ambainis08,EARAC10}.


\subsection{$2 \mapsto 1$ RAC protocol}
\label{sec:2a}
\subsubsection{Classical scheme}
In the classical scenario, Alice's role involves adopting a suitable strategy and encoding a randomly generated 2-bit string into a single bit, which she then sends to Bob. Upon receiving this transmitted bit, the task given to Bob is to extract information about any randomly chosen bit of Alice's original string of 2 bits. This is sought to be obtained from the encoded single bit message with a certain probability of success ($\pr$). Here, it needs to be noted that Alice does not dictate Bob on which of the 2 bits he needs to determine. The protocol can be described on the basis of Alice's \& Bob's chosen encoding-decoding scheme as follows. Say, for input 2-bit strings of the form $X=xx$ ($x \in \{0,1\}$), Alice adopts some appropriate procedure $\hat{A}_{1}$ to encode the single bit message, while for $X=x\bar{x}$ ($x \in \{0,1\}$), she adopts procedure $\hat{A}_{2}$. Correspondingly, Bob employs some decoding process $\hat{B}_{1}$ on this single bit message when tasked to extract the first bit, and he implements $\hat{B}_{2}$ when he requires the second bit. These "procedures" may then, also be viewed as binary-outcome "measurements", with each $\hat{A_{i}}$ and $\hat{B_{j}}$ yielding outcomes in $\{0,1\}$. It is convenient to map these binary values via $0 \leftrightarrow +1$ and $1 \leftrightarrow -1$. So for instance, if Alice’s input string is $01$, she applies $\hat{A}_{2}$ to generate an outcome ${\pm 1}$, which serves as the transmitted single bit message. If Bob is asked to infer the first bit (which is 0 in this case), he applies $\hat{B}_{1}$, if instead he is required to infer the second bit (which is 1), he opts for $\hat{B}_{2}$. The final output is reported in accordance with the prescribed mapping. Therefore, the success of the protocol hinges on Bob's ability to make a precise estimation based on the outcome of his measurement on the single bit information he receives from Alice. In this formulation, all encoding and decoding measurements are dichotomic, yielding values in $\{\pm 1\}$. \vspace{1mm}

In order to optimise the success probability of the protocol, Alice and Bob are permitted to meet beforehand and formulate a \textit{strategy}. This strategy involves a carefully calibrated agreement on encoding and decoding processes between Alice and Bob. For instance, they may agree that Alice will transmit the first bit of her 2-bit string as the encoded message. This would enable Bob to simply reproduce the received bit if he is asked to estimate the first bit of Alice's 2-bit string and to make a random guess if he is required to estimate the second bit. Alternatively, they could employ majority encoding and identity decoding functions, as explained later, in order to retrieve information about the string of 2 bits originally generated. Consequently, Alice's measurement outcomes ($A_i$) will be correlated with Bob's measurement outcomes ($B_j$) based upon their agreed encoding-decoding strategy. \vspace{1mm}

\noindent Having thus formulated the basics of the classical 2-bit scenario, we now proceed to show the existence of a suitably chosen deterministic strategy for which the success probability can attain an optimal value. To begin with, note that the condition of success used in this treatment is given by
\begin{equation} \label{eq:1}
    y = x_{k} \mid X, k \,\,,
\end{equation}
where $X$ is the initial string of 2 bits generated at Alice's end, $x_k$ denotes what Bob is required to estimate, namely, the $k^{\rm th}$ bit of Alice's 2-bit string, and $y$ represents Bob's actual estimation of this $k^{\rm th}$ bit. \vspace{1mm}

\noindent Then, the success probability of the classical protocol ($\pcl$) can be expressed as follows,
\begin{equation} \label{eq:2}
    \pcl = \sum_{X, y, k} V(X, k, y) P_{X} P_{k} P(y \mid X, k) \,\,.
\end{equation}
Here, $V(X, k, y)$ defines a function which outputs 1 (0) subject to satisfying (not satisfying) the condition of success stated in the preceding paragraph, $P_{X}$ is the probability that a particular 2-bit string $X$ is generated, $P_{k}$ represents the probability that Bob is assigned the task of estimating the $k^{\rm th}$ bit, and $P(y \mid X, k)$ defines the probability of Bob making an estimation $y$ of this $k^{\rm th}$ bit of the initially generated string of bits $X$. \vspace{1mm}

\noindent Using the condition given by Eq.~(\ref{eq:1}), $\pcl_2$ can then be obtained as given by (for details see Appendix \ref{A}),
\begin{equation} \label{eq:3}
        \pcl_{2} = \frac{1}{8} \sum_{X, y, k} P(y = x_{k} \mid X, k) \,\,.
\end{equation}
It can then be shown (see App.~\ref{app_optimize}) that this success probability is maximized using a \textit{deterministic} strategy. Consequently, the same upper bound applies to the maximum average success probability achievable using any \textit{randomized} strategy, since every randomized strategy can be expressed as a probability distribution over deterministic ones. On the other hand, in the present work we show that a fundamental bound on the success probability exists irrespective of specific strategies by leveraging a noncontextuality-based Bell-CHSH type inequality.
\vspace{2mm} \\
In the classical framework, Alice's encoding measurements yield outcomes that are defined independent of what outcomes Bob's decoding measurements yield. Under this assumption, the correlations between Alice's and Bob's measurement results can be constrained using an inequality, arising from the notion of noncontextuality of measurement outcomes. To this end, we invoke a Bell-CHSH type inequality formulated within this noncontextual framework~\cite{home1984}. This inequality captures the necessary classicality of any resource underlying the classical $2 \mapsto 1$ RAC protocol, and is therefore directly relevant for establishing any upper bound on $\pcl_{2}$. For our purpose, we use the single particle Bell-type inequality (SPBI) originally derived in~\cite{basu2001}, which takes a form similar to the celebrated Bell-CHSH inequality~\cite{CHSH}. Such a choice is further motivated by the fact that, in the quantum scenario considered subsequently, the protocol is implemented using entanglement within a \textit{single particle} as resource. Then, introducing $\mathscr{C}_2$ to denote the value of the Bell-CHSH type expression for brevity, the form of the SPBI is given by
\begin{equation}  \label{eq:7}
    \left|\left\langle A_1 B_1\right\rangle+\left\langle A_1 B_2\right\rangle+\left\langle A_2 B_1\right\rangle-\left\langle A_2 B_2\right\rangle\right| = \left|\mathscr{C}_2\right| \leq 2 \,\,.
\end{equation}
In this expression, $\hat{A}_{i}$s and $\hat{B}_{j}$s represent two pairs of dynamical variables, such that $\hat{A}_{i}$ and $\hat{B}_{j}$ are co-measurable. Furthermore, each of the individual outcomes $A_1$, $A_2$, $B_1$, and $B_2$ is assumed to take one of two discrete values (\textit{i.e.}, $\pm 1$). It is worth noting that the upper bound given by Eq.~(\ref{eq:7}) on the value of the Bell-CHSH type expression ($\mathscr{C}_2$) is a classical noncontextual bound obtained using the assumption that the values ascribed to $A_i$s are independent of which $B_j$ is measured and what value is ascribed to it. It then follows that $-2 \leq \mathscr{C}_2 \leq 2$. Then, by invoking the assumption of induction~\cite{leggett08} as in the case of the Bell-CHSH inequality, we can interpret the averages $\left\langle A_1 B_1 \right\rangle$, $\left\langle A_1 B_2 \right\rangle, \ldots$ as actual empirically measured quantities. Then, the expression for $\langle A_i B_j \rangle$ can be written in terms of observable probabilities as
\begin{equation} \label{eq:8}
\begin{aligned}
    \langle A_i B_j \rangle &= (-1)(-1)P(11 \mid A_{i}B_{j}) + (-1)(1)P(10 \mid A_{i}B_{j}) \\
    &\quad + (1)(-1)P(01 \mid A_{i}B_{j}) + (1)(1)P(00 \mid A_{i}B_{j}) \,\,.
\end{aligned}
\end{equation}
The complete expression of $\mathscr{C}_2$ then reads as
\small
\begin{equation} \label{eq:9}
    \begin{aligned}
    \mathscr{C}_2 = P(11 \mid A_{1}B_{1}) - P(10 \mid A_{1}B_{1}) - P(01 \mid A_{1}B_{1}) + P(00 \mid A_{1}B_{1})& \\
    + P(11 \mid A_{1}B_{2}) - P(10 \mid A_{1}B_{2}) - P(01 \mid A_{1}B_{2}) + P(00 \mid A_{1}B_{2})& \\
    + P(11 \mid A_{2}B_{1}) - P(10 \mid A_{2}B_{1}) - P(01 \mid A_{2}B_{1}) + P(00 \mid A_{2}B_{1})& \\
    - P(11 \mid A_{2}B_{2}) + P(10 \mid A_{2}B_{2}) + P(01 \mid A_{2}B_{2}) - P(00 \mid A_{2}B_{2})& \,.
    \end{aligned}
\end{equation}
\normalsize
\\
On the other hand, writing explicitly the summation in Eq.~(\ref{eq:3}), the success probability ($\pcl_{2}$) can be expressed as follows,
\small
\begin{equation} \label{eq:10}
\begin{aligned}
    \pcl_{2} = &\frac{1}{8}\Bigl[P\left(y=0 \mid X_{00} B_1\right)+P\left(y=0 \mid X_{00} B_2\right) \\ 
    &\,\,\, +P\left(y=0 \mid X_{01} B_1\right)+P\left(y=1 \mid X_{01} B_2\right) \\
    &\,\,\, +P\left(y=1 \mid X_{10} B_1\right)+P\left(y=0 \mid X_{10} B_2\right) \\
    &\,\,\, +P\left(y=1 \mid X_{11} B_1\right)+P\left(y=1 \mid X_{11} B_2\right) \Bigr] \,\,.
\end{aligned}
\end{equation}
\normalsize
Note that, in the above expression, the terms which do not contribute to success have been implicitly omitted. 
\vspace{2mm} \\
Now, comes the crux of our argument. Note that Bob’s output $y$ corresponds to the outcome $B_j = (-1)^y$, and using the success condition $y = x_k$, is equivalent to $B_j = (-1)^{x_k}$. Then, having obtained Eqs.~(\ref{eq:9}) \& (\ref{eq:10}) in the context of classical $2 \mapsto 1$ RAC protocol, the connection between the values of the \textit{Bell-CHSH type expression} ($\mathscr{C}_2$) and the \textit{success probability} ($\pcl_{2}$) is uncovered in App.~\ref{B}. The upshot is the following relationship,
\begin{equation} \label{eq:11}
    \pcl_{2} = \half \left( 1 + \frac{\mathscr{C}_2}{4} \right) \,\,.
\end{equation}
This interplay can be demonstrated by considering different encoding-decoding strategies for all the possible input bit strings (specifics in App.~\ref{C}). Since the classical (noncontextual) upper bound on $\mathscr{C}_2$ is 2, as stipulated by the SPBI given in Eq.~(\ref{eq:7}), then according to Eq.~(\ref{eq:11}), $\pcl_{2}$ is constrained by the upper limit of $\half \left(1+\half \right) = \frac{3}{4}$. Another important implication of Eq.~(\ref{eq:11}) is that it also provides the lower bound on $\pcl_{2}$. This lower limit is $\half \left(1-\half \right) = \frac{1}{4}$, corresponding to the minimum value of $\mathscr{C}_2 = -2$ (see App.~\ref{C} for more details on this case). Therefore, combining these results, we note that $\frac{1}{4} \leq \pcl_{2} \leq \frac{3}{4}$. This arises directly from the classical notion of noncontextuality, endowing the obtained bounds with complete generality, beyond the confines of any particular strategy.



\subsubsection{Quantum scheme harnessing intraparticle entanglement}
\label{sec:2a2}

\begin{figure}[t!]
\centering{\includegraphics[width=\columnwidth]{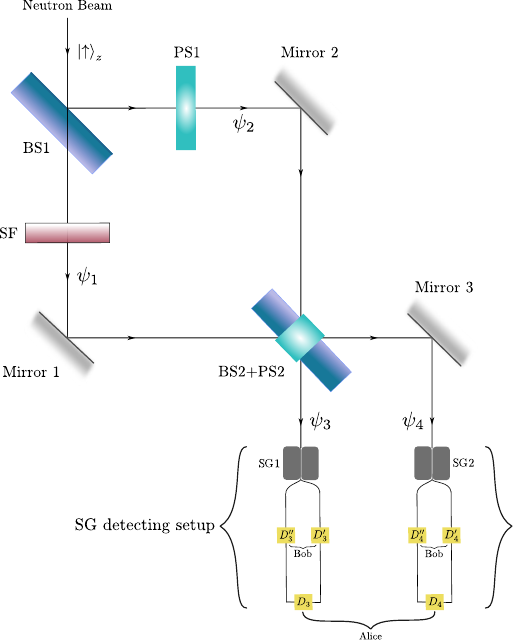}}
\caption{\textbf{\underline{Mach-Zehnder arrangement for the scheme}}: Spin-polarized neutrons passing through beamsplitters (BS1 \& BS2+PS2) are prepared in the intraparticle entangled state of Eq.~(\ref{eq:13}), followed by spin measurements using two Stern-Gerlach devices (SG1 \& SG2). Appropriate settings $(\theta,\phi)$ of BS2+PS2 and SG orientations enable joint measurements of path ($\hat{A_i}$) and spin ($\hat{B_j}$). Detector counts are recorded such that Alice at $D_{3,4}$ measures path (via combined channels) and Bob at $D_{3,4}^{\prime}$, $D_{3,4}^{\prime\prime}$ measures spin.
}
\label{Fig: 1}
\end{figure}

Having established the classical RAC scheme and the corresponding bounds on $\pcl_{2}$, it is natural to examine whether a formulation based on quantum mechanical resources offers any advantage. Moreover, it is of interest to identify the origin of such an advantage. Motivated by these considerations, we now present a quantum version of the RAC protocol that harnesses intraparticle entanglement. We begin our formulation by describing a Mach-Zehnder (MZ) interferometric arrangement (Fig.~\ref{Fig: 1}) that is used to produce an intraparticle path-spin entangled state for the considered protocol. A spin-$\half$ particle (such as a neutron) entering such an interferometric arrangement through a beamsplitter BS1 emerges in mutually exclusive states $\ket{\psi_{1}}$ and $\ket{\psi_{2}}$ corresponding to the transmitted and reflected channels respectively. The states $\ket{\psi_{1}}$ and $\ket{\psi_{2}}$ reunite at a second beamsplitter, BS2, which is used in tandem with a phase-shifting arrangement (PS2), yielding output states $\ket{\psi_{3}}$ and $\ket{\psi_{4}}$, which are detected at $D_3$ and $D_4$ respectively. These spatial states span a two-dimensional Hilbert space $\mathscr{H}_1$, which is \textit{isomorphic} (explained in App.~\ref{app_isomorphism}) to the spin-$\frac{1}{2}$ Hilbert space $\mathscr{H}_2$. Each $\ket{\psi_{i}}$ (for $i \in \{1,2,3,4\}$) is treated as an eigenstate of the path projection operator $P(\psi_i)$ associated with observables that signify the determination of 'which channel' the particle occupies. Here, the detector $D_i$ records the particle's presence (absence) in the channel $\psi_{i}$, such that the outcome of this binary measurement corresponds to the eigenvalue of $P(\psi_{i})$ being +1 (0). It is evident that such a measurement constitutes a coarse-grained position measurement, entailing an exchange of momentum as well. \vspace{1mm}

We now define the path observable $\hat{A} = P(\psi_{3}) - P(\psi_{4})$, whose eigenvalues $\pm 1$ correspond to particle detection in $\psi_{3}$ or $\psi_{4}$ respectively. The expectation value $\langle A \rangle$ can be determined from the relative counts recorded at detectors $D_3$ and $D_4$. More generally, the eigenstates of path observables $\hat{A}_i$ are linear combinations of $\ket{\psi_3}$ and $\ket{\psi_4}$. This is determined by the parameters $\theta$ and $\phi$ of the BS2+PS2 arrangement, whose variation effectively selects different measurement bases, analogous to measuring spin along different orientations with respect to the $z$-axis. Experimentally, such tunability can be achieved, for instance, by employing a parallel-sided plate as a phase shifter to vary $\phi$, and a magnetically saturated Heusler crystal, combined with a rectangular spin rotator, for tuning $\theta$, as in prior experiments with neutrons~\cite{hasegawa03}. Therefore, by appropriately adjusting the parameters $\theta$, $\phi$ of BS2+PS2 and the orientations of the Stern-Gerlach devices, one can implement suitable choices of $\hat{A_i}$ and $\hat{B_j}$ respectively, which together define the quantum strategy by fixing the measurement settings for Alice and Bob. Since path observables $\hat{A}$ within the Hilbert space $\mathscr{H}_1$, in this context, evidently commute with spin variables $\hat{B}$ in the disjoint Hilbert space $\mathscr{H}_2$, this framework enables a test of noncontextuality via the SPBI~\cite{basu2001}. The resulting quantum-mechanical violation is precisely the resource leveraged within our scheme to enhance the protocol's success probability.

\vspace{1mm} \noindent In our quest to realise intraparticle path-spin entanglement, we introduce a spin-flipper (SF) and a phase-shifter (PS1) along channels $\psi_{1}$ and $\psi_{2}$. Detectors $D_3$ and $D_4$ are strategically coupled with SG devices, enabling the measurement of both path $\hat{A}$ and spin $\hat{B}$ observables jointly. This is achieved such that the detector $D_i$, for $i \in \{3,4\}$, registers the combined counts of $D_{i}^{\prime}$ and $D_{i}^{\prime \prime}$, where counts at the unprimed detectors correspond to path measurements $\hat{A}_i$, and those at the primed detectors correspond to spin measurements $\hat{B}_j$. We initiate our experiment with a spin-$\half$ particle with initial spin $\ket{\uparrow}_z$, incident on BS1 with the transmission and the reflection probabilities given by $|a|^{2}$ \& $|b|^{2}$, respectively. The ensuing entangled state is expressed as
\begin{equation}  \label{eq:12}
    \ket{\Psi} = a \left( \ket{\psi_1} \otimes \ket{\downarrow}_z \right) + b e^{i \delta} \left( \ket{\psi_2} \otimes \ket{\uparrow}_z \right) \,\,.
\end{equation}
Here, $\ket{\downarrow}_z$ and $\ket{\uparrow}_z$ signify states corresponding to spin components $\sigma_{z} = -1 \text{ \&} +1$, respectively. The isomorphism between $\mathscr{H}_1$ and $\mathscr{H}_2$ permits us to denote $\ket{\psi_{1}}$ as “up” ($\ket{\uparrow}_p$), and $\ket{\psi_{2}}$ as “down” ($\ket{\downarrow}_p$) states in the 2-dimensional position space. Choosing BS1 to achieve a 50\% reflectivity/transmissivity ratio and adjusting PS1 to set $\delta = \pi$, one can finally obtain a maximally entangled state,
\begin{equation} \label{eq:13}
    \ket{\Psi} = \frac{1}{\sqrt{2}} \left( \ket{\uparrow}_p \otimes \ket{\downarrow}_z - \ket{\downarrow}_p \otimes \ket{\uparrow}_z \right) \,\,.
\end{equation}
This state closely resembles the EPR-Bohm entangled state, but involves both the path and the spin degrees of freedom of a single spin-$\half$ particle. With respect to the state $\ket{\Psi}$ in Eq.~(\ref{eq:13}), we consider measurements of the path observables $\hat{A}_i$ and spin observables $\hat{B}_j$, such that Alice encodes information in one degree of freedom (such as path), while Bob decodes using the other (spin). \vspace{1mm}

While the above discussion has been presented in the context of a neutron interferometric arrangement, the essence of our analysis remains generally applicable for any bosonic or fermionic single system. In particular, an empirical test of our analysis can also be implemented using photonic systems by employing standard optical elements such as beam splitters, half-wave plates, and polarizing beam splitters. In such a setup, one can generate the intraparticle entangled state, $\frac{1}{\sqrt{2}} ( \ket{\uparrow}_p \otimes \ket{H} - \ket{\downarrow}_p \otimes \ket{V} )$, where $\ket{H}$ and $\ket{V}$ denote horizontal and vertical polarization states, respectively. The details and experimental feasibility of realizing such photonic intraparticle entangled states and their various applications have been discussed in~\cite{azzini20}, but the RAC protocol remains so far experimentally unexplored using intraparticle entangled states.
\vspace{2mm} \\
We now seek to study the way in which violation of the SPBI, which is a consequence of the violation of noncontextuality, provides a quantitative measure of the quantum contextual advantage in the protocol. Our approach involves preparing intraparticle entanglement and appropriately choosing measurement bases for Alice and Bob to leverage the resulting correlations for their advantage. Specifically, the entangled state $\ket{\Psi}$ is utilised to establish correlations between Alice's path projection measurements ($\hat{A}_1$ or $\hat{A}_2$) and Bob's spin projection measurements ($\hat{B}_1$ or $\hat{B}_2$). We denote the projectors associated with Alice's and Bob's local projective measurements as ${A}_{i}^{a}$ and $B_{j}^{y}$, respectively. Here, ${A}_{i}^{a}$ ($B_{j}^{y}$) projects to the eigenstate with eigenvalue $a$ ($y$) of Alice's (Bob's) measurement $\hat{A_{i}}$ ($\hat{B_{j}}$). Therefore, $P(y \mid X, k)$ from Eq.~(\ref{eq:4}) may be rewritten as
\begin{equation}  \label{eq:14}
    P(y \mid X, k) = \bra{\Psi} A_{i}^{a} \otimes B_{j}^{y} \ket{\Psi} \,\,.
\end{equation}
Hence, akin to Eq.~(\ref{eq:3}), it is possible to express the success probability in the quantum scenario ($\pqm_{2}$) as
\begin{equation} \label{eq:15}
\begin{split}
    \pqm_{2} &= \frac{1}{8} \sum_{X, y, k} V(X, k, y) \bra{\Psi} A_{i}^{a} \otimes B_{j}^{y} \ket{\Psi} \\
    &= \frac{1}{8} \left[ \sum_{X, j, y} \operatorname{Tr}\left( \rho_X B_{j}^{y} \right) \right] \\
    &= \frac{1}{8} \Bigl[ \operatorname{Tr}\Bigl( \rho_{00} B_{1}^{0} + \rho_{00} B_{2}^{0} + \rho_{11} B_{1}^{1} + \rho_{11} B_{2}^{1} \\
    &\hspace{1.2cm} + \rho_{01} B_{1}^{0} + \rho_{01} B_{2}^{1} + \rho_{10} B_{1}^{1} + \rho_{11} B_{2}^{0} \Bigr) \Bigr] \,\,.
\end{split}
\end{equation}
Here, $\rho_X$ denotes the state suitably prepared by Alice's projective measurement (${A}_{i}^{a}$) on $\ket{\Psi}$, encoding information about the generated 2-bit string, $X$. To this end, note that Alice can encode any of the 4 possible 2-bit strings into the qubit $\rho_X$ by performing a suitable dichotomous measurement $\hat{A}_i$ on the initially prepared state $\ket{\Psi}$ given by Eq.~(\ref{eq:13}). Each such $\rho_X$ therefore represents preparation of an appropriate state for the $2 \mapsto 1$ RAC protocol. Making a note of this, Alice's measurement are selected such that their resulting eigenstates are same as that of the general preparations for the $2 \mapsto 1$ RAC, \textit{i.e.}, 
\begin{equation*}
    \rho_{x x}=\half \left[\mathbb{I}+(-1)^x \hat{A_1} \cdot \vec{\sigma}\right], \quad \& \quad \rho_{x \bar{x}}=\half \left[\mathbb{I}+(-1)^x \hat{A_2} \cdot \vec{\sigma}\right],
\end{equation*}
where, $\hat{A_i}$ for $i \in \{1,2\}$ correspond to the Bloch vectors denoting Alice’s measurement directions, and $\vec{\sigma} = \sigma_{x} \, \hat{i} + \sigma_{y} \, \hat{j} + \sigma_{z} \, \hat{k}$ such that $\sigma_{x/y/z}$ are the Pauli matrices.
\vspace{2mm} \\
Further, we can rewrite Eq.~(\ref{eq:8}) in the following form,
\begin{equation} \label{eq:16}
\begin{aligned}
    \left\langle A_1 B_1\right\rangle &= P(11 \mid A_{1}B_{1}) - P(10 \mid A_{1}B_{1}) \\
    &\quad - P(01 \mid A_{1}B_{1}) + P(00 \mid A_{1}B_{1}) \\
    &= \half \operatorname{Tr}\left[ \rho_{11}B_{1}^{1} - \rho_{11}B_{1}^{0} - \rho_{00}B_{1}^{1} + \rho_{00}B_{1}^{0} \right] \\
    &= \half \operatorname{Tr}\left[ \rho_{11}B_{1}^{1} - \rho_{11}(\mathbb{I} - B_{1}^{1}) - \rho_{00}(\mathbb{I} - B_{1}^{0}) + \rho_{00}B_{1}^{0} \right] \\
    &= \operatorname{Tr}\left[ \rho_{11}B_{1}^{1} + \rho_{00}B_{1}^{0} \right] - 1 \\
    &= 1 - \operatorname{Tr}\left[ \rho_{00}B_{1}^{1} + \rho_{11}B_{1}^{0} \right] \,\,.
\end{aligned}
\end{equation}
Then, using similar such expressions for $\left\langle A_1 B_2 \right\rangle$, $\left\langle A_2 B_1 \right\rangle$, \& $\left\langle A_2 B_2 \right\rangle$, Eq.~(\ref{eq:9}) can rewritten as
\begin{equation} \label{eq:17}
\begin{aligned}
    \mathscr{C}_2 &= \left\langle A_1 B_1\right\rangle+\left\langle A_1 B_2\right\rangle+\left\langle A_2 B_1\right\rangle-\left\langle A_2 B_2\right\rangle \\
    &= \operatorname{Tr}\Bigl[ \rho_{11}B_{1}^{1} + \rho_{00}B_{1}^{0} + \rho_{11}B_{2}^{1} + \rho_{00}B_{2}^{0} \\
    &\qquad + \rho_{10}B_{1}^{1} + \rho_{01}B_{1}^{0} + \rho_{01}B_{2}^{1} + \rho_{10}B_{2}^{0} \Bigr] - 4 \,\,.
\end{aligned}
\end{equation}
Then, upon comparing Eqs.~(\ref{eq:15}) \& (\ref{eq:17}), one notes that the interplay between $\pqm_{2}$ and $\mathscr{C}_2$ is embodied in the relation $\mathscr{C}_{2} = 8\pqm_{2} - 4$. This relation has same structural form to that obtained for the classical protocol \textit{i.e.}, Eq.~(\ref{eq:11}). Thus, the following generalised relation holds for both quantum and classical scenarios of the $2 \mapsto 1$ RAC protocol,
\begin{equation} \label{eq:18}
    \operatorname{P}_{2} = \half \left( 1 + \frac{\mathscr{C}_2}{4} \right) \,\,.
\end{equation}
Then, by employing the appropriate measurement bases as detailed in~\cite{EARAC10}, from Eq.~(\ref{eq:18}) we notice that the \textit{quantum} probability of success $(\pqm_{2})$ exceeds that of failure by the quantity given by $\mathscr{C}_2/4 = 1/\sqrt{2}$. This corresponds to the maximum quantum violation of the SPBI given by Eq.~(\ref{eq:7}) with the quantum mechanical value of $\mathscr{C}_2$ achieving $2\sqrt{2}$. Consequently, the maximum success probability in this scheme that utilizes intraparticle entanglement as resource attains $\pqm_{2} = \half \left(1 + \frac{1}{\sqrt{2}}\right)$, which surpasses the maximal success probability, $\pcl_{2} = \frac{3}{4}$, achievable with classical resources/strategies. We emphasize that the same relation, Eq.~(\ref{eq:18}) holds for both the classical and the quantum versions of the RAC protocol because $\mathscr{C}_{2}$ quantifies the strength of correlations between Alice and Bob. Since the protocol is entirely characterized by these correlations, it is possible to express the success probability, $\operatorname{P}_{2}$ in terms of a Bell-CHSH-type parameter, thereby providing a unified description of both classical and quantum implementations within a single framework. The distinction between classical and quantum resources is thus entirely embedded in the value of $\mathscr{C}_{2}$.
\vspace{2mm} \\
Furthermore, by writing the Bell-type parameter $\mathscr{C}_2$ in terms of the counts registered by the detectors, one can express $\pqm_{2}$ using Eq.~(\ref{eq:18}) explicitly in terms of the experimentally measurable quantities (for details, see App.~\ref{D}). Next, we show how the enhancement of $\pqm_{2}$ over $\pcl_{2}$ is quantitatively commensurate with the violation of Eq.~(\ref{eq:7}). From Eq.~(\ref{eq:18}) note that,
\begin{equation} \label{eq:19}
    \pqm_{2} = \pcl_{2} + \frac{\beta_{2}}{8} \,\,,
\end{equation}
where $\beta_{2} = \mathscr{C}_{2}^{\textsl{qm}} - \mathscr{C}_{2}^{\textsl{cl}}$ is the magnitude of violation of the noncontextuality-based SPBI. It is therefore evident that the enhancement of the success probability in the quantum scenario compared to its classical analogue is quantitatively commensurate with the violation of the Bell-type inequality evidenced in our formulated protocol. Since $\pqm_{2}$ can be written explicitly in terms of the counts at the detectors, Eq.~(\ref{eq:19}) represents an experimentally testable correspondence between the enhanced success probability of the quantum protocol and the quantum mechanical violation of Eq.~(\ref{eq:7}).
\vspace{2mm} \\
Moreover, we notice that when the protocol is executed $N$ times using single particle qubits in the state $\ket{\Psi}$ given by Eq.~(\ref{eq:13}), Alice will perform a $\pi$ rotation on Bob’s qubit for approximately half of them, while the other half will remain unrotated. Here the striking feature is that for each of these subensembles, the measurement statistics maximally violate Eq.~(\ref{eq:7}), which results in $\pqm_{2}$ attaining its upper bound. This perspective on the quantum scenario of the $2 \mapsto 1$ RAC, therefore, illustrates its enhanced performance compared to the classical protocol being related to a Bell-CHSH type noncontextuality test. Our results indicate that in such a setup, $\operatorname{P}_{2} > \frac{3}{4}$ reveals nonclassicality in the form of single particle path-spin measurement contextuality, manifested by utilising intraparticle entanglement, distinct from Kochen-Specker contextuality~\cite{KSpeker68} and KCBS contextuality~\cite{KCBS}. Additionally, see that the value of $\operatorname{P}_{2} = 1$ corresponds to $\mathscr{C}_{2} = 4$, underscoring the interconvertibility between our contextuality-based ”racbox” and the corresponding ”PR box”. A detailed understanding of the equivalence between these supraquantum resources can be found in~\cite{app8}.


\subsection{$3 \mapsto 1$ RAC protocol}
\label{sec:2b}
The 3-bit RAC protocol can be considered as a straightforward extension of its 2-bit counterpart. An extensive discussion of the classical 3-bit RAC scheme is available in~\cite{ambainis08}. Quantum versions of this protocol have also been discussed in~\cite{ambainis08,EARAC10}. Here instead, we focus on finding the correspondence between the success probability of the protocol and the violation of a suitable single particle Bell-type inequality relevant to this protocol. For this purpose, we again use intraparticle entanglement as the resource within the same interferometric setup as in the $2 \mapsto 1$ protocol (Fig.~\ref{Fig: 1}), with the key distinction being the measurement bases chosen by Alice and Bob. As discussed in Sec.~\ref{sec:2a2}, appropriate tuning of the parameters $\theta$, $\phi$ of BS2+PS2 and the orientations of the SG devices enables Alice and Bob to implement measurements in their chosen bases. Harnessing this feature, here too, we employ the intraparticle entangled state $\ket{\Psi}$ of Eq.~(\ref{eq:13}) in our quest to relate the protocol’s success probability enhancement to the violation of noncontextuality. \vspace{1mm}

In our $3 \mapsto 1$ RAC scenario, Alice encodes information about the input string of 3 bits through her measurements $\hat{A_{i}}$ on (say) the path degree of freedom. Bob decodes this information and makes an estimation of a randomly selected $k^{\rm th}$ bit of the initial 3-bit string by using his measurements $\hat{B_{j}}$ on (say) the spin degree of freedom of the same particle. Akin to the quantum scenario of the $2 \mapsto 1$ RAC, Alice and Bob employ specific measurement bases to take advantage of the correlations within the path-spin entangled state. A detailed specification of the appropriate measurement bases can be found in~\cite{EARAC10}.
\vspace{1mm} \\
We begin by noting that analogous to Eq.~(\ref{eq:3}), the success probability ($\operatorname{P}_{3}$) of the $3 \mapsto 1$ RAC protocol is
\begin{equation} \label{eq:20}
    \operatorname{P}_{3} = \frac{1}{24} \left[\sum_{X, i} P\left(y=x_i \mid X, i\right) \right] \,\,,
\end{equation}
where, $X = x_1 x_2 x_3$ represents the originally generated 3-bit string that is to be encoded. \vspace{1mm}

At this stage, we formulate an appropriate single particle measurement noncontextuality-based Bell-type inequality pertinent to this protocol. It is worth recalling here that an inequality of identical form and bound has previously been derived in the context of locality-based Bell inequalities~\cite{gisin}. In the present setting, the corresponding noncontextuality inequality is given by
\begin{equation} \label{eq:21}
\begin{split}
    \left|\mathscr{C}_3\right| &= |\left\langle A_1 B_1\right\rangle + \left\langle A_2 B_1\right\rangle - \left\langle A_3 B_1\right\rangle - \left\langle A_4 B_1\right\rangle \\
    &\quad + \left\langle A_1 B_2\right\rangle - \left\langle A_2 B_2\right\rangle + \left\langle A_3 B_2\right\rangle - \left\langle A_4 B_2\right\rangle \\
    &\quad + \left\langle A_1 B_3\right\rangle - \left\langle A_2 B_3\right\rangle - \left\langle A_3 B_3\right\rangle + \left\langle A_4 B_3\right\rangle| \leq 6 \,\,.
\end{split}
\end{equation}
Then, proceeding analogously to the classical $2 \mapsto 1$ case, by expanding the summation in Eq.~(\ref{eq:20}), expressing $\mathscr{C}_{3}$ in terms of probabilities, and comparing the resulting expressions, we obtain the following relation,
\begin{equation} \label{eq:3classical}
    \pcl_{3} = \half \left( 1 + \frac{\mathscr{C}_3}{12} \right) \,\,.
\end{equation}
Next, the relation given by Eq.~(\ref{eq:20}) can be recast for the quantum scenario of the protocol as
\begin{equation} \label{eq:22}
\begin{split}
    \pqm_{3} &= \frac{1}{24} \sum_{X, y, k} V(X, k, y) \bra{\Psi} A_{i}^{a} \otimes B_{j}^{y} \ket{\Psi} \\
    &= \frac{1}{24} \Bigl[ \operatorname{Tr}\Bigl( \rho_{000}\left(B_{1}^{0}+B_{2}^{0}+B_{3}^{0}\right) + \rho_{001}\left(B_{1}^{0}+B_{2}^{0}+B_{3}^{1}\right) \\
    &\quad\quad + \rho_{010}\left(B_{1}^{0}+B_{2}^{1}+B_{3}^{0}\right) + \rho_{100}\left(B_{1}^{1}+B_{2}^{0}+B_{3}^{0}\right) \\
    &\quad\quad + \rho_{011}\left(B_{1}^{0}+B_{2}^{1}+B_{3}^{1}\right) + \rho_{101}\left(B_{1}^{1}+B_{2}^{0}+B_{3}^{1}\right) \\
    &\quad\quad + \rho_{110}\left(B_{1}^{1}+B_{2}^{1}+B_{3}^{0}\right) + \rho_{111}\left(B_{1}^{1}+B_{2}^{1}+B_{3}^{1}\right) \Bigr) \Bigr] \,\,.
\end{split}
\end{equation}
where, $\rho_{x_1 x_2 x_3}$ denotes the state resulting from Alice's projective measurement $A_{i}^{a}$ on $\ket{\Psi}$, and $B^{y}_{j}$ is the projector corresponding to the eigenstate associated with the outcome $y$ of Bob's measurement $\hat{B_{j}}$ on the state prepared by Alice. \vspace{1mm}

Note that Alice can encode the initial 3-bit string $X$ into the qubit $\rho_X$ by performing a suitable dichotomous measurement $\hat{A}_i$ on the initially prepared state represented by Eq.~(\ref{eq:13}). This corresponds to the general preparations of the following states for the $3 \mapsto 1$ RAC:
\begin{equation*}
\begin{aligned}
    \rho_{x x x}=\half \left[\mathbb{I}+(-1)^x \hat{A_1} \cdot \vec{\sigma}\right], \quad \rho_{x x \bar{x}}=\half \left[\mathbb{I}+(-1)^x \hat{A_2} \cdot \vec{\sigma}\right], \\
    \rho_{x \bar{x} x}=\half \left[\mathbb{I}+(-1)^x \hat{A_3} \cdot \vec{\sigma}\right], \quad \rho_{x \bar{x} \bar{x}}=\half \left[\mathbb{I}+(-1)^x \hat{A_4} \cdot \vec{\sigma}\right].
\end{aligned}
\end{equation*}
where, $\hat{A_i}$ for $i \in \{1,2,3,4\}$ denote the Bloch vectors corresponding to Alice’s measurement directions, and $\vec{\sigma} = \sigma_{x} \, \hat{i} + \sigma_{y} \, \hat{j} + \sigma_{z} \, \hat{k}$. Alice's measurements are chosen such that the resulting eigenstates are the same as that of the preparations specified above. Then, by calculating each of the $\langle A_i B_j \rangle$s explicitly like in the 2-bit case, we note the following general form, $\langle A_i B_j\rangle = \left(-1\right)^{a}\left[ \operatorname{Tr}\left[ \rho^{A_{i}^{a}}B_{j}^{0} + \rho^{A_{i}^{\bar{a}}}B_{j}^{1} \right] - 1 \right]$, where, one chooses $a$ to be 0 (1) for the terms in Eq.~(\ref{eq:21}) accompanied by a positive (negative) sign. Here,
\begin{equation} \label{eq:rho^{A_{i}^{a}}}
\rho^{A_{i}^{a}} = 
\begin{cases}
\rho_{aaa} & \text{\,when\,\,} i = 1 \:, \\
\rho_{aa\bar{a}} & \text{\,when\,\,} i = 2 \:, \\
\rho_{a\bar{a}a} & \text{\,when\,\,} i = 3 \:, \\
\rho_{a\bar{a}\bar{a}} & \text{\,when\,\,} i = 4 \:.
\end{cases}
\end{equation}
This leads to rewriting $\mathscr{C}_3$ as
\begin{equation} \label{eq:23}
    \mathscr{C}_3 = \sum_{i=1}^4 \sum_{j=1}^3 (-1)^{a} \operatorname{Tr}\left[\rho^{A_{i}^{a}} B_j^0 + \rho^{A_{i}^{\bar{a}}} B_j^1 \right] - 12 \,\,.
\end{equation}
Using Eqs.~(\ref{eq:22}) and (\ref{eq:23}), we establish the following connection between the success probability of the quantum protocol ($\pqm_{3}$) and the value of the Bell-type parameter ($\mathscr{C}_{3}$),
\begin{equation} \label{eq:24}
    \pqm_{3} = \half \left( 1 + \frac{\mathscr{C}_3}{12} \right) \,\,.
\end{equation}
Adopting suitable measurement bases given in~\cite{EARAC10}, $\mathscr{C}_3$ achieves a maximum quantum mechanical value of $4\sqrt{3}$. One can easily verify that this corresponds to the maximum quantum violation of the measurement noncontextuality-based inequality given in Eq.~(\ref{eq:21}). Accordingly, from Eq.~(\ref{eq:24}), the success probability of this intraparticle entanglement-based $3 \mapsto 1$ RAC scheme achieves the maximal value of $\pqm_{3} = \half \left( 1 + \frac{1}{\sqrt{3}} \right)$. In contrast, by using the classical noncontextual bound on $\mathscr{C}_3$ given by Eq.~(\ref{eq:21}), the success probability expressed in Eq.~(\ref{eq:3classical}) is constrained to $\pcl_{3} = \frac{3}{4}$. This corresponds to the maximum success probability of the $3 \mapsto 1$ RAC protocol by making use of classical strategies/resources~\cite{ambainis08}. Additionally, as with the 2-bit case, the lower bound on $\pcl_{3}$ for deterministic classical strategies is given by $\half \left(1+\frac{-6}{12}\right) = \frac{1}{4}$. \vspace{2mm}

In light of the structural similarity between Eqs.~(\ref{eq:3classical}) \&~(\ref{eq:24}), one can generalise them to encompass both classical and quantum scenarios of the 3-bit RAC protocol as follows,
\begin{equation} \label{eq:25}
    \operatorname{P}_{3} = \half \left( 1 + \frac{\mathscr{C}_3}{12} \right) \,\,.
\end{equation}
Once again, this is possible because the correlations between Alice and Bob are fully encoded in $\mathscr{C}_{3}$. Consequently, $\mathscr{C}_{3}$ inherently captures information regarding the classical or quantum nature of the resource powering the protocol. \vspace{1mm}

To elucidate the connection between the enhanced success probability of the quantum scenario and the quantum violation of the Bell-type noncontextuality inequality under consideration, we define $\beta_{3} = \mathscr{C}_{3}^{\textsl{qm}} - \mathscr{C}_{3}^{\textsl{cl}}$ as the magnitude of this quantum mechanical violation. From Eq.~(\ref{eq:25}), this yields the following commensurability relation,
\begin{equation} \label{eq:26}
    \pqm_{3} = \pcl_{3} + \frac{\beta_{3}}{24} \,\,.
\end{equation}
Importantly, note that in this case, too, similar to the 2-bit RAC case (App.~\ref{D}), $\pqm_{3}$ can be explicitly written in terms of the counts registered at the detectors and hence Eq.~(\ref{eq:26}) embodies an empirically verifiable consequence of the commensurability between the enhanced success probability of our quantum protocol and the quantum mechanical violation of Eq.~(\ref{eq:21}).


\subsection{$n \mapsto 1$ generalisation}
\label{sec:2c}
In the final part of this work, we generalise our results to an $n$-bit RAC protocol. For this purpose, we emphasize that the bounds we obtain for the success probabilities of our intraparticle entanglement-assisted protocols apply equally well to that of the QRAC protocols, as argued in~\cite{EARAC10}. The basic framework and notable features of a general EARAC protocol are outlined in Sec.~\ref{sec:2}.

\begin{figure*}[t!]
\centering
    \begin{subfigure}{0.26\textwidth}
        \centering
        \includegraphics[width=0.42\linewidth]{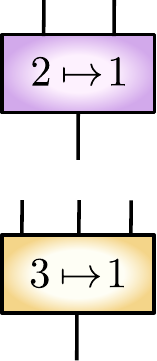}
        \vspace{1.5mm}
        \caption{A schematic illustration of the two subunits: $2 \mapsto 1$ EARAC and $3 \mapsto 1$ EARAC. Each of these subunits essentially represents the setup described in Fig.~\ref{Fig: 1}. The relevant measurement bases are selected appropriately.}
        \label{fig:2(a)}
    \end{subfigure}
    \hfill
    \begin{subfigure}{0.34\textwidth}
        \centering
        \includegraphics[width=0.75\linewidth]{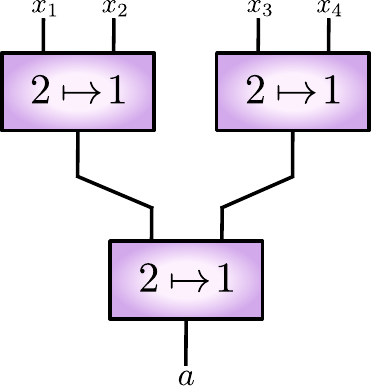}
        \caption{An example of concatenation for the $4 \mapsto 1$ RAC: The probabilities of Bob correctly estimating any $x_k$ is given by Eq.~(\ref{eq:31}) as obtained in~\cite{EARAC10} and is found to be the same for every $k$ in this case because of the same sequence of subunits being used. This probability is evaluated as $\operatorname{P}_{2,0} = \frac{1}{2} \left(1 + \frac{1}{2}\right)$.}
        \label{fig:2(b)}
    \end{subfigure}
    \hfill
    \begin{subfigure}{0.34\textwidth}
        \centering
        \includegraphics[width=0.75\linewidth]{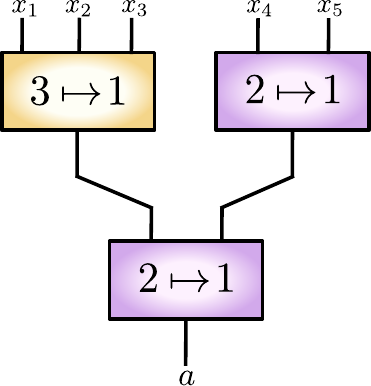}
        \caption{An example of concatenation for the $5 \mapsto 1$ RAC: Here, the probability of Bob correctly estimating $x_4$ is higher than that for $x_3$ because in contrast to $x_3$, only $2 \mapsto 1$ EARAC subunits are used for encoding $x_4$, as shown by the relevant probabilities $\operatorname{P}_{2,0} = \frac{1}{2} \left(1 + \frac{1}{2}\right)$, and $\operatorname{P}_{1,1} = \frac{1}{2} \left(1 + \frac{1}{\sqrt{6}}\right)$.}
        \label{fig:2(c)}
    \end{subfigure}
    \caption{Schematic representation of the concatenation scheme for EARACs}
    \label{fig:2}
\end{figure*}

\vspace{1mm} A key feature of these EARAC protocols is that any $n \mapsto 1$ EARAC (for $n \geq 4$) can be constructed from a combination of the $2 \mapsto 1$ and $3 \mapsto 1$ EARACs through a concatenation scheme (for an elaborate discussion, see~\cite{EARAC10}). In essence, a concatenation scheme entails the combination of multiple 2-bit and 3-bit EARACs by splitting the initial string of $n$ bits into groups of 2-bit and 3-bit ones, encoding them separately by the corresponding EARAC protocols. Alice's classical single bit messages obtained from each such protocol are then used as input bits to be encoded in the next step of the scheme, continuing until Alice has only a single bit left (completing $n \mapsto 1$ encoding). This final encoded bit is then communicated to Bob for decoding. Illustrations of some examples of this scheme are given in Fig.~\ref{fig:2}.

\vspace{1mm} For instance, consider a $4 \mapsto 1$ EARAC protocol which is constructed by using three $2 \mapsto 1$ EARACs (the scheme is illustrated pictorially in Fig.~\ref{fig:2(b)}). The original to-be-encoded 4-bit string is divided into two 2-bit parts, each encoded by a $2 \mapsto 1$ EARAC. The resulting two single bits from these independent encodings are then further encoded by another $2 \mapsto 1$ EARAC, thereby producing the final single-bit message for the $4 \mapsto 1$ EARAC protocol. A crucial consideration for constructing optimal $n$-bit EARACs is that the 2-bit protocol has a higher success probability compared to that of the 3-bit one, and hence the success probability of a concatenated $n \mapsto 1$ EARAC protocol will depend on the composition of the constituent $2 \mapsto 1$ \& $3 \mapsto 1$ EARAC components.

\vspace{1mm} Now, note that a distinctive feature of EARACs enabling concatenation is the use of classical inputs and outputs at each stage, unlike QRACs. In the context of our formulated intraparticle entanglement-based setup, importantly, the counts at the SG detectors corresponding to Alice's measurement in each subunit of the scheme can be directly translated as the classical message/bit to the next stage of the concatenation process. Furthermore, note that by leveraging our intraparticle entanglement-based framework within such a concatenation scheme, the resource cost of the protocol will be significantly reduced. Each subunit (\textit{i.e.}, the 2-bit and 3-bit EARACs) no longer requires two spatially separated entangled particles; instead, a single particle suffices. This reduction in resource usage accumulates over all the steps of the concatenation scheme. This renders our single-particle-based setup particularly well suited for the experimental implementation of concatenated EARAC schemes.

\vspace{2mm} \noindent Having explained above the operational advantage pertaining to the implementability of the intraparticle entanglement-based framework for concatenated EARACs, we now proceed to evaluate the success probability of such a scheme. To this end, note that within the general framework of an $n \mapsto 1$ RAC protocol, the success probability ($\pr$) can be expressed as follows,
\begin{equation} \label{eq:27}
    \pr = \frac{1}{n \hspace{0.5mm} 2^{n}} \left[\sum_{X,k} P\left(y = x_k \mid X, k\right) \right] \,\,,
\end{equation}
where Bob decodes the $k^{\rm th}$ bit of an initial $n$-bit string, $X = x_1 x_2 \ldots x_{n-1} x_n$ to be $y$. \vspace{2mm}

\noindent A comprehensive discussion of the classical $n \mapsto 1$ RAC has been made in~\cite{ambainis08}, where the following expression for success probability of the optimal protocol has been obtained,
\begin{equation} \label{eq:28}
    \pcl_{n} = \half + \frac{1}{2^n}\binom{n-1}{\left\lfloor\frac{n-1}{2}\right\rfloor} \,\,.
\end{equation}
We now invoke a Bell-type inequality that aptly characterises the noncontextuality embodied in the classical $n \mapsto 1$ protocol. Such an inequality can be expressed in the following concise form,
\begin{equation} \label{eq:29}
        \mathscr{C}_n \leq \sum_{r=0}^{\left\lfloor\frac{n-1}{2}\right\rfloor}(n-2 r)\binom{n}{r} \,\,.
\end{equation}
Here, the value of the Bell-CHSH type expression is denoted by $\mathscr{C}_n$, $\left\lfloor \hspace{1mm} \right\rfloor$ denotes the floor function, and $\binom{}{}$ denotes the binomial coefficient. It is worth noting that the above inequality is similar to the general form of Gisin's "elegant Bell inequalities" discussed in~\cite{gisin}.
\vspace{2mm} \\
Then, considering Eqs.~(\ref{eq:28}) \& (\ref{eq:29}), we obtain (for details, see App.~\ref{E}) the following expression for success probability in this case,
\begin{equation} \label{eq:30}
    \pr = \half\left( 1 + \frac{\mathscr{C}_n}{n \hspace{0.5mm} 2^{n-1}} \right) \,\,.
\end{equation}
Building on the discussions in Secs.~\ref{sec:2a} \& \ref{sec:2b}, we note that $\mathscr{C}_{n}$ inherently contains information about the classical or quantum nature of the correlations between the parties in the $n \mapsto 1$ RAC protocol. Consequently, Eq.~(\ref{eq:30}) remains valid in the quantum scenario as well, with $\mathscr{C}_{n}$ taking the value appropriate to the quantum implementation. Thus, having expressed the success probability in terms of the value of the relevant Bell-CHSH-type expression, we now proceed to describe the quantum scenario of the protocol. This involves concatenated 2-bit and 3-bit EARACs, each of which can be realised in the same way as detailed in Secs.~\ref{sec:2a} \& \ref{sec:2b}, respectively. In Fig.~\ref{fig:2(a)}, each "boxed subunit" represents the setup depicted in Fig.~\ref{Fig: 1}, with the counts at the SG detectors being appropriately processed and transmitted as a single bit message to the next step of the concatenation scheme.
\vspace{2mm} \\
Determining the exact success probability for such a concatenated $n$-bit EARAC protocol is straightforward. It has been derived in~\cite{EARAC10}, and is expressed in the following form,
\begin{equation} \label{eq:31}
    \operatorname{P}_{k,j} = \half \left(1+2^{-\frac{k}{2}} 3^{-\frac{j}{2}}\right) \,\,,
\end{equation}
where $k$ \& $j$ denote the the number of times 2-bit and 3-bit EARACs, respectively, are used in encoding the bit in question. However, note that Eq.~(\ref{eq:31}) can be put into a neater form in the special case when the success probabilities for all the individual bits are equal and do not involve a randomization procedure using Shared Randomness (SR)~\cite{EARAC10}. More precisely, such a case occurs only when $n=2^{k}3^{j}$; where $k$ \& $j$ have the same meanings as mentioned above. \vspace{2mm}

Upon equating Eqs.~(\ref{eq:30}) \& (\ref{eq:31}) and using the condition mentioned in the preceding paragraph, the quantum mechanical value of $\mathscr{C}_n$ is obtained as $2^{n-1}\sqrt{n}$. This can be understood in the following manner. Note that, at each stage of the concatenation Alice and Bob share the maximally entangled intraparticle state given by Eq.~(\ref{eq:13}). Alice performs projective measurements on this state as described in Sec.~\ref{sec:2a} (for $2 \mapsto 1$ encoding) \& Sec.~\ref{sec:2b} (for $3 \mapsto 1$ encoding), subsequent to which, Bob measures the state in his chosen bases, as required. Once Alice's projective measurement is implemented, it optimizes Bob's probability of correctly estimating the bit corresponding to his own measurement. In this construction, an initial bit is encoded using $k$ and $j$ instances of individual 2-bit and 3-bit EARACs respectively. Accordingly, the quantum mechanical value of $\mathscr{C}_{n}$ includes the factors of $\left(\sqrt{2}\right)^{k} \left(\sqrt{3}\right)^{j} = \sqrt{n}$ and the number of general preparations for an $n$-bit string. Consequently, $\Cqm = 2^{n-1}\sqrt{n}$. Combining this with the fundamental bound on mutual information in any information-processing task~\cite{Pawlowski2009_inf_causality}, together with the corresponding result for EARACs~\cite{EARAC10}, we obtain a \textit{quantum} bound on the success probability of the protocol. Using Eq.~(\ref{eq:30}), this upper bound is given by
\begin{equation} \label{eq:32}
    \pqm_{n} \leq \half + \frac{1}{2\sqrt{n}} \,\,.
\end{equation}
Interestingly, any $n \mapsto 1$ QRAC with SR is then also bounded by the same success probability~\cite{ambainis08, EARAC10}. Thus, this upper bound applies universally to the success probabilities of both quantum formulations of the RAC protocol. This implies $\Cqm \leq 2^{n-1}\sqrt{n}$. Furthermore, in cases where $n$ does not conform to the form $2^{k}3^{j}$, the parties can employ an $n_{\geq} \mapsto 1$ EARAC, where $n_{\geq}$ is the smallest integer greater than $n$ that adheres to the form $2^{k}3^{j}$~\cite{EARAC10}. This approach will yield,
\begin{equation} \label{eq:33}
    \pqm_{n} \geq \half + \frac{\mathscr{C}_{n_{\geq}}^{qm}}{n_{\geq} \, 2^{n_{\geq}}} \,\,.
\end{equation}
Using Eq.~(\ref{eq:30}), one can now quantitatively characterize the advantage arising from the quantum mechanical violation of the inequality given in Eq.~(\ref{eq:29}). Let $\beta_{n} = \Cqm - \Ccl$ denote the magnitude of this violation. Then, one obtains the relation,
\begin{equation} \label{eq:34}
    \pqm_{n} = \pcl_{n} + \frac{\beta_{n}}{n \, 2^{n}} \,\,.
\end{equation}
Note that in the large $n$ limit, one recovers $\pqm_{n} \approx \pcl_{n} \approx 0.5$ as expected~\cite{ambainis08,EARAC10}. Therefore, Eq.~(\ref{eq:34}) succinctly captures the way in which the quantum advantage in any $n$-bit RAC protocol-enabled quantum information task depends on the amount of quantum path-spin contextuality, evidenced by the magnitude of quantum violation of the relevant Bell-type inequality.


\section{Summary \& Outlook}
\label{sec:3}
Over the last few decades, the quantum information theoretic protocols have largely been based on entanglement between properties of spatially separated particles and have harnessed the nonclassical property of nonlocality (For an up-to-date review, see for example,~\cite{nl_rev}). On the other hand, the possibility of utilizing the quantumness of a single particle by invoking the essentially single particle quantum property of contextuality as an ingredient for the information theoretic tasks has only been studied in recent years~\cite{qkd1,qkd2,qkd3,swap,qrep,qt1,qt2}. \vspace{2mm}

Among the various information theoretic schemes, the Random Access Code (RAC) protocol occupies a significant place, giving rise to wide ranging applications~\cite{app1,netcode,app2,app6,app5,app3,app4,app9,app10,app11,app7,app8,app13,app12,app_teleport_2026}. In the context of the RAC protocols, apart from the studies based on interparticle entanglement, the only ones based on a single particle employ either temporal correlations by harnessing quantum violations of the Leggett-Garg type inequalities~\cite{tempRAC}, or preparation contextuality towards implementing the RAC-like parity oblivious multiplexing task~\cite{spek09,ghorai}. To enlarge this domain of applications, we analyse in this paper for the RAC protocols, a novel application of a form of quantum measurement contextuality arising essentially from path-spin entanglement of a single particle. A key ingredient of this work is the analysis of the success probabilities in the intraparticle entanglement resourced $2$-bit, $3$-bit, and $n$-bit RAC protocols in light of the Bell-type inequalities derived from noncontextuality of single particle path-spin measurement outcomes. Importantly, our scheme is formulated in an experimentally amenable way in terms of the setups already used in the prior studies of intraparticle entanglement using photon \& neutron interferometries. An upshot of this study is the uncovering of the feature that the success probabilities in the quantum scenarios of the RAC protocols exceed their classical counterparts by amounts which are quantitatively commensurate with the quantum violations of the relevant Bell-type single particle noncontextuality inequalities. Ramifications of this finding in light of some of the recent works~\cite{RAC_Tavakoli15,RAC_Hameedi17,RAC_Liabotro17,RAC_Carmeli20,kanj23,sahadas23,RAC_Gatti23,RAC_Ambainis24,RAC_Farkas25} concerning fundamental aspects and applications of RAC protocols call for further probing. \vspace{2mm}

Here it needs to be pointed out that the exact relationship between nonlocality and quantum advantage in the RAC protocols powered by interparticle entanglement remains unexplored. Motivated by the present analysis, one may anticipate that a similar quantitative correspondence, analogous to that established here between contextuality and quantum advantage, also holds in that setting. This expectation is reinforced by the fact that the descriptions of intraparticle and interparticle entanglement are mathematically equivalent, and hence the intraparticle entanglement-resourced quantum advantages for the RAC protocols discussed here are quantitatively equivalent to those achievable using interparticle entanglement as resource. However, in contrast to interparticle entanglement, which requires maintaining coherence between the states of sufficiently separated particles, our protocol offers the operational advantage that the generation and preservation of intraparticle entanglement involving a single particle is less demanding. In this connection, one may note the studies~\cite{saha16,adc} concerning the robustness of intraparticle entanglement against decoherence and dephasing. Notably,~\cite{adc} also uncovers the intriguing possibility of noise-aided generation of intraparticle entanglement using even an initially unentangled state as resource. In light of such studies, it will therefore be useful to comprehensively investigate the robustness of the intraparticle entanglement-based RAC protocols formulated in this paper, taking into account the possible presence of different types of decoherence/damping which may occur in actual experimental implementations, as well as by considering non-idealness of the devices used in the experiments. \vspace{2mm}

It has not escaped our notice that the path-spin contextuality enabled maximum quantum probabilities of success in our formulated intraparticle entanglement-based $2$-bit and $3$-bit RAC protocols have, respectively, turned out to be exactly the same as that in the $2$-bit and $3$-bit RAC-like Parity-Oblivious Multiplexing tasks hinging on preparation contextuality~\cite{spek09}, despite the classical optimal success probabilities for the two schemes being different in the $3$-bit case. Unravelling the deeper origin of such striking agreement should be quite instructive. Another promising direction is to take cue from the studies~\cite{hameedi,pan19,saha19b} probing the efficacy of the different forms of contextuality in various quantum communication protocols. This can lead to useful exploration of the questions such as whether the notion of single particle spatial-spin contextuality can have practical applications in some other suitable context (information transfer or processing tasks/communication games) by outperforming the use of other forms of contextuality. Studies towards this end are being pursued to be presented in a sequel paper.
\vspace{5mm} \newline
\noindent \textbf{Acknowledgements:} NS would like to acknowledge the support received through the Department of Science and Technology (DST) under its INSPIRE-SHE programme. NS also thanks IISER, Mohali for facilitating this collaboration. DH dedicates this paper to late Shyamal Sengupta on his birth centenary with whom the ideas of intraparticle entanglement and single particle spatial-spin noncontextuality-based Bell-type inequality were first conceived. DH acknowledges stimulating discussions with Urbasi Sinha and her group on various aspects of intraparticle entanglement. DH also acknowledges support from NASI Senior Scientist Fellowship in initiating this collaboration at Bose Institute, Kolkata. In completing this work, DH acknowledges support as distinguished visitor to the Light and Matter Physics group, RRI, Bangalore under the National Quantum Mission of the DST.


\appendix
\begin{widetext}

\section{Obtaining simplified expression of success probability from its general form}
\label{A}
\noindent From Eq.~(\ref{eq:2}) we have,
\begin{equation} \label{eq:A1}
    \pcl = \sum_{X, y, k} V(X, k, y) P_{X} P_{k} P(y \mid X, k) \,\,,
\end{equation}
where,
\begin{equation}
    V(X, k, y)= \begin{cases}1 & \text { if } y = x_{k} \mid X, k  \\ 0 & \text { otherwise }\end{cases}
\end{equation}
Thus, the function $V(X,y)$ is an indicator of the \textsl{success frequency} of the protocol. Note that, we have invoked the condition of success, $y = x_{k} \mid X, k$, as in Eq.~(\ref{eq:1}) for the purpose of denoting such success frequency. \vspace{2mm}

\noindent Furthermore, in the context of the $n \mapsto 1$ RAC protocol, the other factors in the RHS of Eq.~(\ref{eq:A1}) assume the following explicit forms. Since the original string of $n$ bits is generated \textsl{randomly} at Alice's end, therefore $P_{X} = \frac{1}{2^{n}}$. Moreover, Alice and Bob do not know in advance which bit of the sequence needs to be estimated. As a result, any of the initial bits can be chosen \textsl{randomly} for Bob to estimate. Then, $P_{k} = \frac{1}{n}$. \vspace{2mm}

\noindent Eq.~(\ref{eq:A1}) is then calculated as,
\begin{equation} \label{eq:prob_gen_app}
        \pcl_{n} = \sum_{X, y, k} V(X, k, y) P_{X} P_{k} P(y \mid X, k) =\frac{1}{n\,2^{n}} \sum_{X, y, k} V(X, k, y) P(y \mid X, k) = \frac{1}{n\,2^{n}} \sum_{X, y, k} P(y = x_{k} \mid X, k) \,\,.
\end{equation}


\section{Optimality of classical success probability}
\label{app_optimize}

\noindent Since Alice and Bob a priori agree upon the encoding-decoding strategy to be adopted, the probability $P(y \mid X, k)$ is contingent upon such an adopted strategy. Here, we introduce $m \in M$ as a variable (independent of the input bit-string $X$ and the measurements $\hat{A}_{i}$ and $\hat{B}_{j}$) for suitably labeling their strategy. Then, for appropriately defined deterministic strategies $\mathcal{S}_{m}(X, k)$, it can be shown that,
\begin{equation}  \label{eq:4}
\begin{aligned}
    P(y \mid X, k) &= \sum_{m} P_{M}(m) \,\, P(y \mid m, X, k) \\
    &= \sum_{m} P_{M}(m) \,\, \delta_{\mathcal{S}_{m}(X, k), y} \,\,.
\end{aligned}
\end{equation}
Then, for the $2 \mapsto 1$ scenario, using Eqs.~(\ref{eq:prob_gen_app}) \& (\ref{eq:4}), the following relation is obtained,
\begin{equation}  \label{eq:5}
    \begin{aligned}
        \pcl_2 &= \frac{1}{8} \sum_{X, y, k} V(X, k, y) P(y \mid X, k) \\
        &= \sum_{m} P_{M}(m) \left[ \frac{1}{8} \sum_{X, k} \delta_{\mathcal{S}_{m}(X, k), x_{k}} \right] \\
        &= \sum_{m} P_{M}(m) \, \pcl_{2}\{m\} \\
        &\leq \pcl_{2}\{m^{*}\} \,\,.
    \end{aligned}
\end{equation}
Here, the inner sum $\pcl_{2}\{m\}$ is interpreted as the success probability corresponding to a deterministic strategy denoted by $m$. Note that while going from the equality to the inequality in Eq.~(\ref{eq:5}), we have introduced $m^{*}$, where $m^{*} \in M$ represents the strategy for which $\pcl_{2}\{m\}$ is maximized. The inequality then implies that the success probability is maximized using a \textit{deterministic} strategy. Example of such a deterministic strategy has been shown in terms of \textit{majority encoding} and \textit{identity decoding} functions in~\cite{ambainis08}. The optimal classical success probability can then be demonstrated to achieve a value of $\pcl_2 = \frac{3}{4}$. One can also extend this observation to obtain the maximum average success probability for any randomized strategy making use of the fact that any randomized strategy can be represented as a probability distribution of deterministic strategies.


\section{Relating $\pcl_{2}$ with $\mathscr{C}_2$}
\label{B}

\noindent We do a case-by-case analysis of Eq.~(\ref{eq:3}),
\be
\pcl_{2} = \frac{1}{8} \sum_{X, y, k} P(y = x_{k} \mid X, k) \,.
\ee

\noindent \underline{For $X=00$}: \\
From the condition of success in Eq.~(\ref{eq:1}), we must have $y=x_{k}=0$. Then, $B_{j}=+1$ for success. Since Alice uses $A_{1}$ to encode (as described in the main paper), we have the following,
\begin{equation} \label{eq:X00_B1}
\begin{aligned}
    P\left( y=0 \mid X_{00} B_{1} \right) &= P\left( A_{1}=+1 , B_{1}=+1 \right) + P\left( A_{1}=-1 , B_{1}=+1 \right) \\
    &= P\left( A_1 B_1 = +1 \right) + P\left( A_1 B_1 = -1 \right) \,,
\end{aligned}
\end{equation}
and
\begin{equation} \label{eq:X00_B2}
\begin{aligned}
    P\left( y=0 \mid X_{00} B_{2} \right) &= P\left( A_{1}=+1 , B_{2}=+1 \right) + P\left( A_{1}=-1 , B_{2}=+1 \right) \\
    &= P\left( A_1 B_2 = +1 \right) + P\left( A_1 B_2 = -1 \right) \,.
\end{aligned}
\end{equation}

\noindent \underline{For $X=01$}: \\
Here, success requires:
\begin{equation}
    y = \begin{cases}0 & \text{ for } k=1  \implies B_{1}=+1 \\ 1 & \text{ for } k=2 \implies B_{2}=-1 \end{cases}
\end{equation}
Alice uses $A_{2}$ to encode. Therefore,
\begin{equation} \label{eq:X01_B1}
\begin{aligned}
    P\left( y=0 \mid X_{01} B_{1} \right) &= P\left( A_{2}=+1 , B_{1}=+1 \right) + P\left( A_{2}=-1 , B_{1}=+1 \right) \\
    &= P\left( A_2 B_1 = +1 \right) + P\left( A_2 B_1 = -1 \right) \,,
\end{aligned}
\end{equation}
and
\begin{equation} \label{eq:X01_B2}
\begin{aligned}
    P\left( y=1 \mid X_{01} B_{2} \right) &= P\left( A_{2}=+1 , B_{2}=-1 \right) + P\left( A_{2}=-1 , B_{2}=-1 \right) \\
    &= P\left( A_2 B_2 = -1 \right) + P\left( A_2 B_2 = +1 \right) \,.
\end{aligned}
\end{equation}

\noindent \underline{For $X=10$}: \\
As an extension of the logic in previous cases, here we obtain
\begin{equation} \label{eq:X10}
\begin{aligned}
    P\left( y=1 \mid X_{10} B_{1} \right) &= P\left( A_2 B_1 = +1 \right) + P\left( A_2 B_1 = -1 \right) \,, \\
    P\left( y=0 \mid X_{10} B_{2} \right) &= P\left( A_2 B_2 = +1 \right) + P\left( A_2 B_2 = -1 \right) \,.
\end{aligned}
\end{equation}

\noindent \underline{For $X=11$}: \\
Similarly,
\begin{equation} \label{eq:X11}
\begin{aligned}
    P\left( y=1 \mid X_{11} B_{1} \right) &= P\left( A_1 B_1 = +1 \right) + P\left( A_1 B_1 = -1 \right) \,, \\
    P\left( y=1 \mid X_{11} B_{2} \right) &= P\left( A_1 B_2 = +1 \right) + P\left( A_1 B_2 = -1 \right) \,.
\end{aligned}
\end{equation}

\noindent Next, note that
\begin{equation} \label{eq:AiBj}
\begin{aligned}
    \langle A_i B_j \rangle &= P(11 \mid A_{i}B_{j}) - P(10 \mid A_{i}B_{j}) - P(01 \mid A_{i}B_{j}) + P(00 \mid A_{i}B_{j}) \\
    &= P_{11} - P_{10} - P_{01} +P_{00}
\end{aligned}
\end{equation}
Then, $P\left( A_i B_j = +1\right) = P_{11} + P_{00}$, and $P\left( A_i B_j = -1\right) = P_{10} + P_{01}$. Also, $P_{11} + P_{10} + P_{01} +P_{00} = 1$. \\

\noindent Combining the above information, it is straightforward to obtain
\begin{equation} \label{eq:<AiBj> to P(AiBj)}
    P\left( A_i B_j = \pm 1 \right) = \frac{1 \pm \langle A_i B_j \rangle}{2} \,\,.
\end{equation}

\noindent Based on choices of deterministic strategies between Alice and Bob, there will be a constraint such that $A_{i} B_{j} = (-1)^{f(X, k)}$, where $f(X, k)$ is a binary-valued function suitably defined for every strategy. Therefore each term, $P(y=x_k \mid X, k)$ reduces to either $P\left( A_i B_j = +1 \right)$ or $P\left( A_i B_j = -1 \right)$. Of course for a "random" strategy where Alice's does a random encoding (randomly send +1/-1) and/or Bob makes a random guess, there will be no such cancellation, but factoring in a normalization factor of $\half$ within each $P(y=x_k \mid X, k)$ and on adding everything up, all the $\langle A_{i} B_{j} \rangle$ terms cancel and $\pcl_{2}$ sums up 0.5. A general expression for the success probability of the classical $2 \mapsto 1$ scheme then stands as,
\begin{equation} \label{eq:general_pcl_2}
\begin{aligned}
    \pcl_{2} &= \frac{1}{8} \sum_{X, k} \frac{1 + (-1)^{f(X,k)} \langle A_i B_j \rangle}{2} \\
    &= \half + \frac{1}{16} \sum_{X, k} (-1)^{f(X,k)} \langle A_i B_j \rangle \,\,.
\end{aligned}
\end{equation}
However, for optimal deterministic strategies such as \textit{majority encoding} \& \textit{identity decoding}, and others~\cite{ambainis08}, the function $f(X,k)$ cannot be chosen such that all constraints $A_i B_j = (-1)^{f(X,k)}$ are simultaneously satisfied for all $(X,k)$. At most three of the four corresponding correlation terms can be aligned, while one must be anti-aligned. This leads to $\sum_{X,k} (-1)^{f(X,k)} \langle A_i B_j \rangle = 2\left( \langle A_1 B_1 \rangle + \langle A_1 B_2 \rangle + \langle A_2 B_1 \rangle - \langle A_2 B_2 \rangle \right)$. Thus,
\begin{equation}
    \pcl_{2} = \half \left( 1+\frac{\mathscr{C}_2}{4} \right) \,,
\end{equation}
where, $\mathscr{C}_{2}$ is the Bell-CHSH type variable arising from the notion of noncontextuality as defined in Eq.~(\ref{eq:7}). \vspace{1mm}

\noindent Here we make an important remark: The Bell-CHSH-type structure is not imposed \textit{a priori}. It arises from the incompatibility inherent in classical deterministic RAC strategies, which cannot simultaneously satisfy all four correlation (success) constraints required for perfect retrieval of both bits. The best one can do is satisfy 3 out of 4 constraints simultaneously, yielding the maximum classical success probability of $\pcl_{2} = 0.75$.


\section{Examples: Interplay between $\pcl_{2}$ and $\mathscr{C}_2$}
\label{C}

\noindent Consider the following explicit form of $\mathscr{C}_2$, which is expressed in Eq.~(\ref{eq:9}) as
\begin{equation}
    \begin{aligned}
    \mathscr{C}_2 &= P(11 \mid A_{1}B_{1}) - P(10 \mid A_{1}B_{1}) - P(01 \mid A_{1}B_{1}) + P(00 \mid A_{1}B_{1}) \\
    &\quad + P(11 \mid A_{1}B_{2}) - P(10 \mid A_{1}B_{2}) - P(01 \mid A_{1}B_{2}) + P(00 \mid A_{1}B_{2}) \\
    &\quad + P(11 \mid A_{2}B_{1}) - P(10 \mid A_{2}B_{1}) - P(01 \mid A_{2}B_{1}) + P(00 \mid A_{2}B_{1}) \\
    &\quad - P(11 \mid A_{2}B_{2}) + P(10 \mid A_{2}B_{2}) + P(01 \mid A_{2}B_{2}) - P(00 \mid A_{2}B_{2}) \,\,.
    \end{aligned}
\end{equation}
and the following expanded complete expression of $\pcl_{2}$, given by Eq.~(\ref{eq:10}) as
\begin{equation}
\begin{aligned}
    \pcl_{2} = \frac{1}{8}&\Bigl[P\left(y=0 \mid X_{00} B_1\right) + P\left(y=0 \mid X_{00} B_2\right) + P\left(y=0 \mid X_{01} B_1\right) + P\left(y=1 \mid X_{01} B_2\right) \\
    &+ P\left(y=1 \mid X_{10} B_1\right) + P\left(y=0 \mid X_{10} B_2\right) + P \left(y=1 \mid X_{11} B_1\right) + P\left(y=1 \mid X_{11} B_2\right) \Bigr] \,\,.
\end{aligned}
\end{equation}
Now, consider a strategy where Alice always sends the first bit of the bit string to Bob as the communicated message. Bob then reproduces this bit when asked to estimate the first bit of the original string and makes a random guess for the second bit. In this strategy, $\mathscr{C}_2$ reaches a value of 2 according to Eq.~(\ref{eq:9}), and $\pcl_{2}$ achieves a value of $\frac{3}{4}$ for each of the 2-bit strings (00, 01, 10, and 11), resulting in an average success probability of $\frac{3}{4}$ using Eq.~(\ref{eq:10}). \\

Next, consider another strategy where Alice always sends the "majority" bit (1 if the number of ones in the original bit string is $\geq n/2=1$, 0 otherwise) to Bob. Bob then reproduces the received bit when asked to estimate any bit in the original string. Here, $\mathscr{C}_2$ again reaches a value of 2 according to Eq.~(\ref{eq:9}), and $\pcl_{2}$ achieves a value of 1 for the bit strings 00 and 11, and $\frac{1}{2}$ for the bit strings 01 and 10. Then, according to Eq.~(\ref{eq:10}) an average success probability of $\frac{3}{4}$ is attained. \\

Consider the strategy where Alice always sends to Bob the "majority" bit of the bit string as the communicated message and Bob reproduces the conjugate of this received bit when asked to estimate any bit in the original bit string. Interestingly, $\mathscr{C}_2$ will attain the minimum value of -2 in this description, according to Eq.~(\ref{eq:9}). Moreover, note that $\pcl_{2}$ will attain a value of $0$ for each of the bit strings 00 \& 11 and it will attain the value of $\frac{1}{2}$ for each of the bit strings 01 \& 10. Then, from Eq.~(\ref{eq:10}) the average success probability is $\frac{1}{4}$. \\

Further, consider a strategy where Bob makes a random guess irrespective of Alice's communicated message. As can be expected in this case, $\mathscr{C}_2$ achieves the value 0 according to Eq.~(\ref{eq:9}). Additionally, note that $\pcl_{2}$ will attain a value of $\frac{1}{2}$ for all possible 2-bit strings, \textit{i.e.}, 00, 01, 10, \& 11. This implies that the average success probability will be $\frac{1}{2}$ according to Eq.~(\ref{eq:10}).
\\ \\
The above analysis involving various encoding-decoding measurement pairs reveals the interplay between $\pcl_2$ and $\mathscr{C}_{2}$ as consistently embodied in the relationship obtained in App.~\ref{B},
\begin{equation} \label{eq:C3}
    \pcl_{2} = \half \left( 1 + \frac{\mathscr{C}_2}{4} \right) \,\,.
\end{equation}
Here an important observation is that the "information" about various strategies is now encapsulated within $\mathscr{C}_2$, and hence, studying its bounds suffices when evaluating the optimal success probability of the protocol. Therefore, the correlations between Alice's and Bob's measurements provide a quantitative measure of success and any strategy that maximises the value of $\mathscr{C}_2$ will also maximise the success probability. In this sense, the result given by Eq.~(\ref{eq:C3}) integrates all encoding-decoding processes between Alice and Bob in a classical/noncontextual scenario.


\section{Isomorphism between the two-dimensional path Hilbert space $\mathscr{H}_1$ and the spin-$\half$ Hilbert space $\mathscr{H}_2$}
\label{app_isomorphism}
\noindent "Which-way"/Interference type measurements can be described in terms of projection operators constructed from the states $\ket{\psi_1}$ and $\ket{\psi_2}$, where the orthonormal basis states $\ket{\psi_1}$ and $\ket{\psi_2}$ span a two-dimensional complex Hilbert space ($\mathscr{H}_1$). The algebra of such projection operators (dichotomic observables) is \textit{isomorphic} with the algebra of Pauli spin operators (for example, see~\cite{takesaki1979}). \\

\noindent We define,
\begin{equation}
\begin{aligned}
P_1 = \ket{\psi_1}\bra{\psi_1} \quad &, \quad P_2 = \ket{\psi_2}\bra{\psi_2}\, , \\
P_{12} = \ket{\psi_1}\bra{\psi_2} \quad &, \quad P_{21} = \ket{\psi_2}\bra{\psi_1}\, , \\
P_3 = \half \ket{\psi_1 + \psi_2}\bra{\psi_1 + \psi_2} \quad &\& \quad P_4 = \half \ket{\psi_1 - \psi_2}\bra{\psi_1 - \psi_2}\, .
\end{aligned}
\end{equation}
Consider a double-slit type experiment with an input state which is a linear combination of $\ket{\psi_1}$ and $\ket{\psi_2}$. The operator corresponding to "which way ($\ket{\psi_1}$ or $\ket{\psi_2}$) measurement" has eigenstates $\ket{\psi_1}$, $\ket{\psi_2}$ with corresponding eigenvalues +1 and -1, respectively. $P_1 - P_2$ is the operator representing such "which way measurement". The operators $P_3$ and $P_4$ represent "interference measurements" corresponding to $\ket{\psi_1 + \psi_2}$ and $\ket{\psi_1 - \psi_2}$, respectively. \vspace{2mm}

\noindent Isomorphism between the algebra of projection operators and the algebra of $2 \times 2$ complex matrices spanned as a linear space by the Pauli matrices $\sigma_x$, $\sigma_y$, $\sigma_z$, and the Identity $\mathbb{I}$ can be expressed explicitly as,
\begin{equation}
    \ket{\psi_1} = \binom{1}{0}, \quad \& \quad \ket{\psi_2} = \binom{0}{1} \hspace{1cm}\text{(for example)}
\end{equation}
One can easily check that
\begin{equation}
\begin{aligned}
    P_1 = \half (\mathbb{I} + \sigma_z) \quad &, \quad P_2 = \half (\mathbb{I} - \sigma_z)\, , \\
    P_{12} = \half (\sigma_x + i\sigma_y) \quad &, \quad P_{21} = \half (\sigma_x - i\sigma_y)\, , \\
    P_1 + P_2 = \mathbb{I} \, , \quad P_1 - P_2 = \sigma_z \hspace{2mm} &, \hspace{2mm} P_{12} + P_{21} = \sigma_x \, , \quad P_{12} - P_{21} = i\sigma_y \,.
\end{aligned}
\end{equation}
Now, if an input linear combination of $\ket{\psi_1}$ and $\ket{\psi_2}$ is transformed into that of $\ket{\psi_3}$ and $\ket{\psi_4}$ (say, by using a beam splitter with reflectivity $\sin\theta$ and transmittance $\cos\theta$)
\begin{equation}
    \ket{\psi_3} = \sin\theta \, \ket{\psi_1} + \cos\theta \, \ket{\psi_2} \quad \& \quad \ket{\psi_4} = \cos\theta \, \ket{\psi_1} - \sin\theta \, \ket{\psi_2}
\end{equation}
where,
\begin{equation}
\begin{aligned}
    P(\psi_3) &= \begin{pmatrix} (1-\cos2\theta)/2 & (\sin2\theta)/2 \\ (\sin2\theta)/2 & (1+\cos2\theta)/2 \end{pmatrix} \,, \\
    P(\psi_4) &= \begin{pmatrix} (1+\cos2\theta)/2 & (-\sin2\theta)/2 \\ (-\sin2\theta)/2 & (1-\cos2\theta)/2 \end{pmatrix} \,.
\end{aligned}
\end{equation}
Then, "which-way" detections pertaining to $\ket{\psi_3}$ and $\ket{\psi_4}$ are described by the observable
\begin{equation}
    P(\psi_3) - P(\psi_4) = \begin{pmatrix} -\cos2\theta & \sin2\theta \\ \sin2\theta & \cos2\theta \end{pmatrix} = (\sin2\theta) \, \sigma_x - (\cos2\theta) \,\sigma_z = \vec{\sigma} \cdot \vec{a} \,\,.
\end{equation}
where, $\, \vec{a} = \sin2\theta \, \hat{i} - \cos2\theta \, \hat{k}$. \\

\noindent This shows how the observables (defined in terms of projection operators) corresponding to "which-way" detection counts can be measured by changing the beam splitter parameter $\theta$ such that their isomorphism to the spin-$\half$ Hilbert space $\mathscr{H}_2$ ensures correspondence with measuring spin components along various directions.


\section{Expressing $\pqm_{2}$ in terms of the experimentally measurable counts registered at the detectors}
\label{D}
\noindent The explicit correspondence between actual measurements in terms of the counts registered at the detectors and the quantities occurring on the LHS of Eq.~(\ref{eq:7}) is given in~\cite{basu2001} as
\begin{equation}
    \left\langle A_1 B_1\right\rangle 
    =\left(N_3-N_4\right)\left[\left(N_3^{\prime}-N_3^{\prime \prime}\right)+\left(N_4^{\prime}-N_4^{\prime \prime}\right)\right] \,\,.
\end{equation}
where the pair $A_1 B_1$ is considered and the registered counts are denoted by $N_3, N_3^{\prime}$, \& $N_3^{\prime \prime}$  (for Detectors $D_3, D_3^{\prime}$, \& $D_3^{\prime \prime}$, respectively), and $N_4$, $N_4^{\prime}$ \& $N_4^{\prime \prime}$ (for Detectors $D_4$, $D_4^{\prime}$ \& $D_4^{\prime \prime}$, respectively). \vspace{2mm}

We utilize this result but adopt a different notation for convenience. We denote the counts registered at the specified detectors for a pair of measurements $A_i B_j$ as $N_{ij}, N_{ij}^{\prime}$, \& $N_{ij}^{\prime \prime}$  (for Detectors $D_3, D_3^{\prime}$, \& $D_3^{\prime \prime}$, respectively), and $M_{ij}$, $M_{ij}^{\prime}$ \& $M_{ij}^{\prime \prime}$ (for Detectors $D_4, D_4^{\prime}$, \& $D_4^{\prime \prime}$, respectively). Then, we must have
\begin{equation}
    \langle A_i B_j\rangle = \left(N_{ij}-M_{ij}\right)\left[\left(N_{ij}^{\prime}-N_{ij}^{\prime \prime}\right)+\left(M_{ij}^{\prime}-M_{ij}^{\prime \prime}\right)\right] \,\,.
\end{equation}
Consequently, Eq.~(\ref{eq:7}) is recast in the following form,
\begin{equation} \label{eq:D3}
    \mathscr{C}_2 = \sum_{i,j = 1}^{2} (-1)^{r} \hspace{0.5mm} \langle A_i B_j\rangle = \sum_{i,j = 1}^{2} (-1)^{r} \hspace{0.5mm} \left[\left(N_{ij}-M_{ij}\right)\left[\left(N_{ij}^{\prime}-N_{ij}^{\prime \prime}\right)+\left(M_{ij}^{\prime}-M_{ij}^{\prime \prime}\right)\right]\right] \,\,.
\end{equation}
where, $r = 0$ for $(i,j) \in \{(1,1),(1,2),(2,1)\}$, and $r = 1$ for $(i,j) \in \{(2,2)\}$. Then, by representing $\mathscr{C}_2$ as given by Eq.~(\ref{eq:D3}) in terms of the detector counts, $\pr$ can be explicitly expressed in terms of the experimentally measurable quantities using Eq.~(\ref{eq:18}) as follows,
\begin{equation}
    \pqm_{2} = \half + \frac{1}{8} \sum_{i,j = 1}^{2} (-1)^{r} \hspace{0.5mm} \left[\left(N_{ij}-M_{ij}\right)\left[\left(N_{ij}^{\prime}-N_{ij}^{\prime \prime}\right)+\left(M_{ij}^{\prime}-M_{ij}^{\prime \prime}\right)\right]\right] \,\,.
\end{equation}


\section{Relating success probability of the $n \mapsto 1$ RAC protocol with the value of the relevant Bell-type expression}
\label{E}
\noindent Consider the bound on Eq.~(\ref{eq:29}) which has the form,
\begin{equation}
    \mathscr{C}_n \leq \sum_{r=0}^{\left\lfloor\frac{n-1}{2}\right\rfloor}(n-2 r)\binom{n}{r} \,\,.
\end{equation}
Let $k = \left\lfloor\frac{n-1}{2}\right\rfloor$ for simplicity. We then express the above summation as follows,
\begin{equation}
    \begin{aligned}
        \sum_{r=0}^{k}(n-2 r)\binom{n}{r} &= \sum_{r=0}^{k} \frac{n!}{(n-r)! r!} (n-r-r) \\
        &= \sum_{r=0}^{k} n\left( \frac{(n-1)!}{(n-1-r)!r!}-\frac{(n-1)!}{\{(n-1)-(r-1)\}!(r-1)!} \right) \\
        &= \sum_{r=0}^{k} n\left(\binom{n-1}r - \binom{n-1}{r-1}\right) \,\,.
    \end{aligned}
\end{equation}
A drastically simplifying cancelling of terms in this "telescoping sum" would then yield:
\begin{equation} \label{eq:E3}
    \begin{aligned}
        &\sum_{r=0}^{k}(n-2 r)\binom{n}{r} = n \hspace{1mm} \binom{n-1}{k} \\
        &\phantom{{}\hspace{4mm}{}} \Rightarrow \mathscr{C}_n \hspace{0.5mm} \leq \hspace{0.5mm} n \hspace{1mm} \binom{n-1}{\left\lfloor\frac{n-1}{2}\right\rfloor}
    \end{aligned}
\end{equation}
Subsequently, substituting Eqs.~(\ref{eq:E3}) \& (\ref{eq:28}) in the expression $\pr = \half + f(n)\mathscr{C}_n$ (taking cue from our earlier discussions in Secs.~\ref{sec:2a}, \ref{sec:2b} \& App.~\ref{B}), the following expression is obtained,
\begin{equation}
    \pr = \half + \frac{\mathscr{C}_n}{n \, 2^{n}} \,\,.
\end{equation}

\end{widetext}



\bibliographystyle{apsrev4-2}
\bibliography{ref}






\end{document}